\documentclass[preprint,12pt]{aastex}
\usepackage{epsfig}


\def\rd{Di\thinspace Stefano} 
\def\bi{binary} 
\def\wdf{white dwarf}
\def\ce{common envelope}
\def\rl{Roche lobe}
\def\m-s{main-sequence}

\def\m-t{mass-transfer}

\def\pr{progenitor}
\def\t1{Type Ia supernova}
\def\s-d{single-degenerate}
\def\d-d{double-degenerate}
\def\dd{double degenerate}
\def\sd{single degenerate}

\begin{document}

\title{
The Progenitors of Type Ia Supernovae:\\ II.~Are they Double-Degenerate Binaries?\\ 
The Symbiotic Channel
}
%: LUMINOUS, BUT NOT SUPERSOFT} 

\author{R. Di\thinspace Stefano
}
\affil{
Harvard-Smithsonian Center for Astrophysics,
60 Garden St..
Cambridge, MA 02138}

\begin{abstract}

In order for a \wdf\ to achieve the Chandrasekhar mass, $M_C$, and explode
as a Type~Ia supernova (SNIa), it must
interact with another star, either
accreting matter from or merging with it.
The failure to
identify the class or classes of binaries which produce SNeIa 
is the long-standing ``\pr\ problem''. Its solution is required if we
are to utilize the full potential of SNeIa to elucidate
basic cosmological and physical principles.
%This paper considers a range of binary models, linking the
%physical characteristics of each to its potentially observable signatures.
%We thereby provide a guide to the discovery of the \pr s 
%of SNeIa.  

In single-degenerate models, a \wdf\ accretes and burns matter
at high rates. Nuclear-burning \wdf s (NBWDs) with mass close to
$M_C$ are hot and
luminous, potentially detectable as supersoft x-ray sources (SSSs).
In previous work we showed that
$> 90-99\%$ of the required number of \pr s do not
appear as SSSs during most of the crucial phase of mass increase.

The obvious implication might be that double-degenerate binaries
form the main class of progenitors. We show in this paper, however,  that
many binaries that later become double-degenerates
must pass through a long-lived
NBWD phase during which they
 are potentially
detectable as SSSs.
The paucity of SSSs is therefore not a strong argument in
favor of \d-d\ models. 
Those NBWDs that are the progenitors of double-degenerate binaries
are likely to appear as symbiotic binaries for intervals $> 10^6$~years.
In fact, symbiotic pre-double-degenerates should be common, whether or
not the white dwarfs eventually produce Type~Ia supernovae.  

The key to solving the
Type~Ia \pr\ problem lies in understanding the appearance of NBWDs.
Most of them do not appear as SSSs most of the time. We therefore consider the
evolution of NBWDs 
to address the question of what their appearance may be and how we can 
hope to detect them.

\end{abstract}

\section{Introduction}

\t1e have been used to map the expansion history of the Universe. The results
have been exciting, indicating epochs of deceleration and
acceleration, and suggesting the presence of dark energy
(see, e.g.,
Riess et al.\, 2007; Kuznetsova et al.\, 2008).
Unfortunately, we have not yet identified the astronomical
systems that produce these distinctive explosions.
 (See Kotak 2009 and Branch
et al.\, 1995 for reviews.) Until we do, it will be
impossible to understand or quantify the systematic uncertainties and to
optimize the further use of \t1e to explore physics and
cosmology.
 The \t1\ \pr\ problem is therefore considered to be
one of the key outstanding questions in astronomy
today.

We know that the explosions occur when a \wdf\ gains mass from a
binary companion. Indications from both theory and observation
are that the supernova is triggered when the \wdf\ reaches the
Chandrasekhar mass, $M_C$ (Mazzali et al.\, 2007). 
What we don't know are the
characteristics of the binary.
Is the donor on the main sequence, evolved, or degenerate? 
Whatever the nature of the donor, it must be able to contribute enough mass
to the \wdf\ to allow it to transition from its starting mass to $M_C.$

In single-degenerate binaries,
the rate of mass transfer to a \wdf\ from a non-degenerate 
donor must be high enough that
matter can be burned 
in either a quasisteady way or else during recurrent novae,
thereby eliminating opportunities for more explosive nuclear burning that
can reduce the mass of the \wdf\ (Iben 1982; Nomoto 1982; Fujimoto 1982). 
That is, 
the \wdf s that reach $M_C$ must process
accreting material; they are nuclear-burning \wdf s (NBWDs)
for long intervals. They are therefore potentially detectable as
hot, luminous  supersoft x-ray sources (SSSs) during the crucial epoch when
the \wdf 's mass is increasing\footnote{The known SSSs typically have 
$30\, {\rm eV} < k\, T < 100\, {\rm eV}$ and 
$10^{36} {\rm erg~s}^{-1} < L_X < 10^{38} {\rm erg~s}^{-1}.$
NBWDs with mass near $M_C$ have surface temperatures and luminosities 
at the top end of these ranges. 
[See Figure 1 of the 
companion paper (Di\thinspace Stefano 2010).]}.    
Some bright SSSs may be progenitors of Type~Ia
supernovae (Rappaport et al.\, 1994; \rd\ \& Rappaport 1994). 
Nevertheless, the companion paper (\rd\ 2010; see also
\rd\ et al.\, 2010 and \rd\ 2007) 
shows conclusively that the majority of
the progenitors do not appear as bright SSSs during intervals long
enough ($\sim 10^5$~yrs) to allow quasisteady burning of the necessary
amounts of accreting matter.
For both spiral and elliptical galaxies,
the discrepancy is at least an order of magnitude, perhaps as much as two orders of magnitude.
In addition, we found that existing data already place restrictions on
sub-Chandrasekhar
models. These restrictions may be tightened as additional
exposures with {\it Chandra} and {\it XMM-Newton} are taken and more data are analyzed.

The most obvious interpretation of the mismatch is that it rules out
single-degenerate models. 
In fact, a weaker measure of the mismatch was recently derived for six
early-type populations, and  was used to argue that single degenerates
can produce no more than $5\%$ of the \t1e\ in early-type galaxies
(Gilfanov \& Bogd{\'a}n 2010).  
If single-degenerates are ruled out, then the alternative would appear
to be double-degenerate models in which two carbon-oxygen (C-O) \wdf s
execute a close orbit. In order for the \wdf s to come to interact
in a Hubble time, they must have had an opportunity to spiral toward
each other in a common envelope.   
This paper explores the epoch prior to the common envelope.
In \S 2 we find that,  
immediately before the common envelope phase that produces a close
double-degenerate,  
an epoch of nuclear burning on an accreting white dwarf is expected.
In \S 3 we predict the numbers of NBWDs required if the \d-d\ channel
is the main route to \t1e. We then 
compare these numbers with the numbers of SSSs
detected in external galaxies, and find a large mismatch. In \S 4 we 
discuss the significance and implications of the mismatch.    
Section 5 focuses on the symbiotic nature of pre-double-degenerate binaries,
and discusses the prospects for using the distinctive symbiotic phase to
 test double-degenerate models for SNIa \pr s.  
Our conclusions are presented in \S 6. 

The bottom line is that, for neither young nor old populations can the
absence of SSSs be interpreted as evidence for the absence of NBWDs.
If the photospheres of NBWDs are large, soft x-rays may not be emitted.
In fact photospheric adjustments in known SSSs seem to occur (see. e.g.,
Greiner \& \rd\ 2002).
 In addition, local mass associated with the system, 
such as winds, can absorb radiation from the \wdf. In fact,
the very binaries most likely
to produce \t1e must eject significant winds if they are to survive    
(\rd\ et al. 1997; \rd\ \& Nelson 1996; \rd\ 1996).

\section{Double Degenerates and Nuclear Burning}  

\subsection{Overview}

In order for two white dwarfs to come close enough to each other to
either exchange mass or
merge in a Hubble time, they must be
separated by a distance no larger than a few $R_\sun$.
This requires a prior common envelope phase, initiated when
a giant with a well-formed
core fills its Roche lobe, typically at a point when it is more massive
than its WD companion.

Below we consider binaries about to enter a common envelope
phase and emerge as double-degenerates with total mass
greater than the Chandrasekhar mass. Even before describing the
calculations, however, we can explain why, prior to the 
common envelope, the white dwarf is likely to burn some of the matter
impinging on it. 
This is because the rate of mass infall, 
$\dot M_{in},$ produced by winds from a giant,  
is comparable to what is needed for quasisteady nuclear burning.
Most of the mass lost by a giant is ejected at rates, $\dot M_{wind},$
ranging from $10^{-7} M_\odot$~yr$^{-1}$ to $10^{-5} M_\odot$~yr$^{-1}$.
In cases in which the giant eventually fills its Roche lobe, the
fraction $f$ of the winds that impinge upon the \wdf\ will start
at just a few percent, and reach roughly $1/4-1/3$ just before the giant
fills its Roche lobe. Thus, the infall rate is likely to be 
in the approximate range 
$\sim 10^{-8} M_\odot$~yr$^{-1}- 10^{-6} M_\odot$~yr$^{-1}.$     
These values are near or above the rate required for quasisteady nuclear
burning, shown as the black curves in Figure 1. For infall rates
between the two curves, matter can be burned more-or-less as it accretes.
For infall rates just below the steady burning regime, 
matter
can accumulate over decades and burn during recurrent novae. The
explosions are weak enough that much of the accreted matter can be
retained. For lower values of the infall rate, classical novae that
blow away most or all of the accumulated matter are expected.
For infall rates above the steady burning regime, not all of the 
incoming mass can be burned. The excess matter could
be ejected and/or accumulate in an envelope around the \wdf . 

\subsection{Generating Double Degenerates and their Predecessors}

Consider a binary in which neither star is massive enough to
produce a core-collapse supernova, and in which the star that was initially the
most massive, star ``1'', has already evolved.   
Let $M_{1,wd}$ represent the mass of the
first-formed \wdf , $M_{2,ce}$ represent the 
total mass 
of the giant just at the point when it fills its Roche lobe, and
$M_{2,wd}$ the mass of its core at the same time. We consider as
potential progenitors of \t1e those binaries in which  the white dwarfs
are C-O \wdf s, with  
%\begin{equation} 
$M_{1,wd}+M_{2,wd}>M_C$ 
%\end{equation}
We use a uniform distribution to select the value of $M_{1,wd}$ to 
lie in the range $0.55\, M_\odot - 1.35\, M_\odot,$ and then the value of 
$M_{2,wd}$ to   
lie in the range $0.45\, M_\odot - M_{1,wd}$.  

The following situation triggers a \ce : 
when star ``2'' fills its Roche lobe, the effect of
mass transfer is to further shrink the Roche lobe, while the star itself
is unable to shrink. In order for this to occur, 
%\begin{equation}
$M_{2,ce} > \eta\, M_{1,wd}.$  
%\end{equation}
Typically, $\eta>1.$ The value of $\eta$ for each individual system depends
on the amount of mass ejected from the system, the angular momentum
carried by ejected mass, and on the mass ratio. 
Keeping the stability criterion in mind, 
we select the values of $M_{2,ce}$ from a uniform distribution:
$M_{1,wd} < M_{2,ce} <  7.6\, M_\odot.$
The upper limit was chosen to allow for the fact that the 
initial mass of the secondary, $M_2,$ is larger than 
$M_{2,ce}$. In addition, the initial mass of the primary is larger than
$M_2.$ We took the upper limit on
$M_1$ to be $8.4,$ roughly corresponding to the most massive star that
becomes a \wdf .

Prior to filling its Roche lobe, star ``2'' 
was emitting a wind. 
We have modeled the wind using a Reimer's-type law,
modified so that it satisfies the following conditions. First,
although there may be modest mass loss during the main-sequence
and subgiant phase, significant mass loss starts only
 when the core mass
reaches a critical value $c_0.$ The Reimer's form then
ensures that $\dot M_{wind}$ increases with time, more dramatically
as the core mass reaches its final value. 
The second condition we impose is that, for stars
evolving in isolation, the integrated mass lost through
winds
is $M_2-M_{2,wd{0}},$ where $M_{2,wd{0}}$ is the mass
a star of initial mass $M_2$ would produce, were it to evolve in
isolation. The values of $M_{2,wd{0}}$ and $M_2$  
are related to each other through 
an initial-mass/final-mass relationship: 
$M_f=0.123\, M_o + 0.358.$
(See Kalirai et al. 2008; Catal{\'a}n et al. 2009 ; Dobbie et al. 2006;
Williams 2007; Weideman \& Koester 1983.) 
We assume that, prior to the point at which the donor filled its
Roche lobe, the decrease
in the giant's mass can be modeled with the analytic form derived by
integrating over the mass lost through winds.

When the giant has a companion, a fraction of the mass it loses
comes under the gravitational influence of the companion.
The geometry of the binary and of the winds, and the wind
 speed, determine how much mass can be captured by the companion. When the
giant
is close enough to its companion that it is about to fill its Roche lobe,
the winds are partially focused. We use the expression
\begin{equation} 
\dot M_{in}=\frac{1}{4} \, \dot M_{winds}\Bigg(\frac{R_2(t)}{R_{RL}(t)} \Bigg)  
\end{equation}
$\dot M_{in}$ represents the rate of mass infall to the \wdf , and 
$R_2(t)$ and $R_{RL}(t)$ are the instantaneous values of the physical
radius and Roche lobe radius of star ``2''.  

In the top panel of Figure~1, $\dot M_{in}$ versus $M_{1,wd}$ 
is shown for systems in which the giant is about to fill its Roche lobe.
Green (red) points correspond to binaries in which the combined \wdf\ 
masses sum to more (less) than $M_C.$ 
Plotted as black curves are $\dot M_{min}$ 
and $\dot M_{max}$ as a function of the
accretor mass.  
These are, respectively, the minimum value of $\dot M_{in}$ for
which quasisteady nuclear burning can occur and the maximum value
of $\dot M_{in}$ for which all of the incoming mass can be burned
as it accretes.
 
The striking feature of this plot is that, for many systems, the 
rate of mass infall at the time of Roche-lobe filling
 is in or near the steady-burning regime. Furthermore, this
would be the case under a wide range of assumptions about the infall rate.
That is, many points on this plot would fall within or 
straddle the steady-burning regime
even if a larger or smaller fraction of the giant's winds
were captured by the \wdf,  or even if the winds were emitted by star ``2'' 
at a somewhat higher or lower rate,
or even if the values of $\dot M_{min}$ and 
$\dot M_{max}$ were to differ somewhat from those shown in the figure.
It is therefore a 
robust result that, just before the common envelope phase, mass
is infalling on many of the first-formed \wdf s  at rates compatible with quasi-steady
nuclear burning.

We next evolve each binary 
represented in the top panel of Figure~1  
backward in time, until the point at which the giant's 
core mass is $c_0.$ 
To compute the evolution we must model the fraction $\beta$ 
of infalling mass
that can be retained by the  
\wdf .  
The 
value of $\beta$ 
depends on how the rate of infall compares with the minimum and
maximum rates compatible with steady nuclear burning, 
$\dot M_{min}$ and $\dot M_{max}.$ 
For $\dot M_{in} < 1/3\, \dot M_{min},$ we assume
that no mass is retained.   For rates of infall is within $1/3$ of
$\dot M_{min},$ we take $\beta=0.4.$ In the steady-burning
regime we use $\beta = 0.8.$ For $\dot M_{in}> \dot M_{max},$
we use $\beta= 0.8\, \dot M_{max}/\dot M_{in}$     
With this prescription we can compute the initial mass,
$M_{1,wd}(0)$  of the first-formed \wdf , and compare it with the
mass $M_{1,wd}(f)$ at the time star ``2'' fills its \rl .  

The results are shown in the bottom panel of Figure~1.
For the systems in our simulation, more than $0.05\, M_\odot$ 
($0.15\, M_\odot$) was
gained by $50\%$ ($37\%$) of the \wdf s in
binaries that would eventually go on to become \dd s with 
$M_{tot} > M_C.$ The distributions of properties among the binaries in our
simulation is unlikely to mirror the distributions found in nature.
Nevertheless, our results show that mass gain by the first-formed
\wdf s in \d-d\ binaries may be common and can often be significant. 
This has a potentially important implication for the rate of \t1e 
that could be produced by \d-d\ binaries, since
mass gain by the first-formed \wdf\ can increase the numbers of
\d-d\ binaries in which the total \wdf\ mass exceeds $M_C.$ 

For the work in
this paper we focus on the 
connection between nuclear burning and detectability.
 Since nuclear burning 
is common among pre-double degenerates, these  
binaries must be bright, and are therefore potentially detectable
during an extended interval just prior to the \ce . 
Near the surface of the \wdf\  nuclear-burning 
produces temperatures and luminosities characteristic of
SSSs.

The pre-double-degenerates in which the total \wdf\ mass will
exceed $M_C$ are not the most common pre-double-degenerates.  
It is therefore important to also 
consider those binaries for
which the total \wdf\ mass is smaller than $M_C.$
These are shown as the red points  
in the bottom panel of Figure~1.
While these binaries
are presumably not good candidates for \t1\ \pr s, they too can experience
long epochs during which nuclear burning and SSS-like
behavior occur.  
In these systems, the first-formed \wdf s 
tend 
to gain less mass. In our simulation, 
only $18\%$ gain more than $0.05\, M_\odot,$ 
$5\%$ gain more than $0.12\, M_\odot,$ and $2\%$ gain more than $0.15\%$.
Nevertheless, because more such binaries are expected, they
can comprise a significant component of bright nuclear-burning \wdf s that
have the potential to be detected as SSSs.

\section{Number of Nuclear-Burning Pre-Double-Degenerates} 

\subsection{Predictions}

We have found that, prior to the common envelope producing the \dd ,
there is a time interval, $\tau_{acc},$ during which the 
values of $\dot M_{in}$ are compatible with quasisteady nuclear
burning and mass retention. Let $\Delta M$ represent the
mass gained during this interval.
\begin{equation}
\Big({{\Delta\, M}\over{0.1\, M_\odot}}\Big) = 
\Big({{\tau_{acc}}\over{{{1}} \times 10^6\, yrs}}\Big)\, 
\Big({{{\beta\, \dot M_{in}}}\over{1\times 10^{-7} M_\odot\, yr^{-1}}}\Big) 
\end{equation} 
The value of $\beta\, \dot M_{in}$ represents an average during
the epoch when the WD's mass is changing.

We now compute the number of NBWDs expected
if \d-d\ binaries are the primary class of \t1\ \pr s.
 This discussion mirrors that  
in \S 2 of \rd\ 2010, which computed the number of NBWDs expected
if single-degenerate binaries are the primary class of \t1\ \pr s. 
The rate of SNe Ia in galaxies is roughly $0.3$
per century per $10^{10} L_\odot$ in blue luminosity
(Cappellaro et al.\, 1993; Turatto et al.\, 1994; 
Dilday et al.\, 2008; Graham et al.\, 2008; Kuznetsova et al.\, 2008;
Poznanski et al. 2007; Panagia et al.\, 2007).
In a galaxy with blue luminosity $L_B,$ the number, $N_{acc}$, of
pre-double-degenerate \t1\ \pr s actively accreting, with masses
within
 $0.1\, M_\odot$
of the value they will have when contact is established is  
\begin{equation}
N_{acc}=
3000 \,  
\Bigg({{\Delta\, M}\over{0.1\, M_\odot}}\Bigg)  
\Bigg({{1\times 10^{-7} M_\odot\, yr^{-1}}\over{\beta \, \dot M_{in}}}\Bigg)  
\Bigg({{L_B}\over{10^{10} L_\odot}}\Bigg).
\end{equation}
If \dd s are the principle \pr s of \t1e, then we expect that galaxies such
as the Milky Way, M31, and other large spirals contain thousands of
actively accreting NBWDs.
 Note that this counts {\it only} the pre-double-degenerates
that could eventually become \t1e. In addition, there are many systems
very similar to the supernova \pr s undergoing the same type of evolution, even though
the total \wdf\ mass will be smaller than  $M_C$. The numbers of NBWDs in galaxies
also includes those in which there will not be a common envelope producing
a close \d-d\ \bi.

Elliptical galaxies house older stellar populations and may therefore
have smaller values of $L_B,$  even though their total stellar mass
may be larger. The rates of \t1e in early-type galaxies suggest that
they should contain $\sim 1/3$ as many NBWDs associated with
pre-double-degenerates that are \pr s of \t1e.  
There should therefore be at least several hundred NBWDs in
each elliptical if the \d-d\ channel is the primary route to \t1e.

It is worth comparing the results above to the parallel results for \sd s 
(\rd\ 2010). The number of NBWDs needed to produce the measured rate of
\t1e via the \d-d\ channel may be {\it larger} then the number needed
for the \s-d\ channel. The reason is that, even though comparatively
little mass may be gained by the \wdf s in pre-double-degenerates,
the \wdf s are less massive and can burn incoming matter when the infall rate
is lower. The interval during which the \wdf\ can burn incoming material can
therefore be longer. This can produce a larger number of active nuclear burners
at one time. There are other differences as well, which we will
discuss below.  

\subsection{Observations}

Because there is a link between nuclear-burning white dwarfs and SSSs,
it is important to compare the numbers 
of NBWDs needed to produce the total rate of \t1e with the 
 numbers of SSSs observed.
The advent of {\it Chandra} and {\it XMM-Newton} made it possible to
detect and identify SSSs in external galaxies at least as 
far from us as the Virgo cluster. 
Because extragalactic SSSs provide little flux, individual
sources may provide fewer than $50-100$ counts with {\it Chandra}.
With {\it XMM-Newton} the count rate is higher, but the number of
background counts is also higher. 
It was therefore 
important to develop clear and reliable ways to identify SSSs in external galaxies.
A small set of nearby galaxies, M31,  M101, M83, M51, M104, NGC4472, and
NGC4697  served as testbeds for algorithms to identify SSSs
(Di\thinspace Stefano
et al.\, 2003a, 2003b, 2004a, 2004b).
When applying and testing our algorithms, 
we found that, in addition to SSSs, whose spectra cuts off at roughly
$1$~keV, there are comparably bright 
X-ray sources that are also soft, but which cut-off
at somewhat higher energy, roughly $2$~keV. These were dubbed quasisoft
sources (QSSs). (See Di~Stefano \& Kong 2004; 
Di~Stefano et al. 2004a; Di~Stefano et al. 2004b.)   

Table 1 of  \rd\ 2010 lists the numbers of SSSs
in (M101, M83, M51, M104, NGC4472, NGC4697). The numbers of SSSs
 are, respectively, (42, 28, 15, 5, 5, 4). 
M101, M83, and M51 are late-type type galaxies.
 The dominant stellar populations in
the bulge-dominated spiral M104, and the ellipticals NGC4472, and NGC4697
are likely to be older. 
For galaxies of all types, the numbers of SSSs are roughly two orders of
magnitude smaller than the numbers of NBWDs required if the double-degenerate
channel is the primary route to \t1e.  

In fact, the true discrepancy may be larger, 
because not every SSS observed in these galaxies is
likely to be a pre-double-degenerate \t1\ \pr . Some of the SSSs, for 
example, have luminosities that are too high to be NBWDs. Others are
classical nova (Pietsch et al.\, 2005; Henze et al.\, 2009), 
with accretion rates not in the band expected (\S 2). 
Others are engaged in stable mass transfer and will not produce
a common envelope.
The true mismatch is therefore likely to be larger than two orders of magnitude.

We might ask if it is possible that the NBWDs in the pre-double-degenerate
\t1\ \pr s could exhibit a hard component in addition to the soft radiation 
associated with nuclear burning. If so, they might appear as QSSs.
The numbers of QSSs detected in (M101, M83, M51, M104, NGC4472, NGC4697) 
are, respectively, (21, 26, 21, 17, 22, 15). The populations
of QSSs are far too small to make up the difference. In fact, the
numbers of x-ray sources that are not either SSS  or QSS
 in these galaxies are, respectively
(24, 74, 56, 100, 184, 72). Thus, there would still be a short fall by an
order of magnitude, even if {\sl all} of the  
X-ray sources were pre-double-degenerate
\t1\ \pr s.\footnote{ 
Of course, most of the bright non-SSS and non-QSS sources we detect in other 
galaxies are likely to be accreting neutron stars or black holes, just as is
the case in the Local Group.}  

The mismatch discovered through the study of the 6 galaxies mentioned above
holds as well for M31, and for all $383$ external galaxies observed
with {\it Chandra} and studied by Liu (2008). Other investigations of 
X-ray sources in 
external galaxies also give results that are consistent with there being a relative;y
small number of very soft sources. See, e.g.,
Sarazin et al.(2001);
Swartz et al.(2002);
Pence et al.(2001);
Jenkins et al.(2005);
Fabbiano et al.(2003). 

\section{The Significance of the Mismatch}

\subsection{Is The Mismatch  Real?}

In assessing the significance of 
this mismatch, it is important to 
address the question of whether the extragalactic 
NBWDs in pre-\d-d\ binaries are bright and hot
enough to be detected as SSSs. 
We have already addressed this question for the NBWDs in
\s-d\ \pr s of \t1e (\rd, 2007; \rd\ et al.\, 2009; \rd\ 2010). 
In \s-d\ models, the \wdf s must have mass close to $M_C$ prior
to explosion. As they gain the mass needed to bring them to the limit, they
will be the brightest and hottest NBWDs, with luminosities near the Eddington
limit and effective values of $k\, T$ over $\sim 80$~eV. 
In several nearby galaxies, x-ray observations 
conducted to date would have been able to
provide a complete census of such NBWDs with
$N_H$ in the range of a few times $10^{21}$~cm$^{-1}.$ 
The mismatch is therefore  
highly significant, and has been known for some time.

The accreting \wdf s in pre-\d-ds will generally not have masses near $M_C.$
In general though, we expect that the first \wdf\ to form will be the more massive
of the two \wdf s that eventually merge. Its mass must therefore be larger than
$0.7\, M_\odot,$ and is likely to be larger than $0.8-0.9\, M_\odot.$
Although they  will not be as hot and bright as more massive \wdf s, 
NBWDs in this mass range can be detected in external galaxies. Figure~2 of \rd\ 2010 shows that
the count rate expected from a \wdf\ of $0.9\, M_\odot$ in M31 would allow it to 
be 
detected by {\it Chandra}, even with an $N_H$ of a few times 
$10^{21}$~cm$^{-2}$. White dwarfs of $0.8\, M_\odot$ would be
detected in M31 with $N_H\sim 10^{21}$~cm$^{-2}$.
 Several M31 fields have been well studied at soft
x-ray wavelengths with {\it Chandra}, 
most notably the Bulge, which has the highest
density of soft sources. While {\it XMM-Newton} may not be 
able to resolve all of the sources near the
nucleus, it has provided the advantage of wide-area coverage.
With deep 
surveys of the body of M31, 
{\it XMM-Newton} should have discovered all of the NBWDs with masses 
above $0.8\, M_\odot.$  The combined results from
{\it Chandra}  (\rd\ et al. 2004) and {\it XMM-Newton} (Orio 2006)
show that the number of SSSs detectable at any given time  
is more than an order of magnitude smaller than the 
number predicted by Equation 3.
In addition, deep observations by {\it Chandra} of the face-on galaxy M101 have
been taken ($> 1$ megasecond); the analysis of Liu (2008) finds no evidence of
large-enough numbers of SSSs or QSSs.  

Finally, although gas in the Galactic disk prevents us 
from detecting many SSSs,
Figure~2 shows that the {\it ROSAT All-Sky Survey (RASS)} would have detected
many Galactic NBWDs with masses of $0.9\, M_\odot$ or $0.8\, M_\odot$, if they
emit soft x-rays with the expected luminosities and temperatures. 
Excluding regions with high $N_H$, including star-forming regions and
the direction toward the Galactic center,
NBWDs with the
temperature and luminosity expected for a 
\wdf\ with $0.9\, M_\odot$  ($0.8\, M_\odot$)  
could be identified if they lie within $\sim 6$ ($\sim 3$)~kpc. 
These regions are likely to include at least a few percent of the Galactic
pre-double-degenerates. If, therefore, the \d-d\ channel is the main route 
to \t1e, we expect that the   {\it (RASS)} would have identified dozens of
SSSs. Only a handful were identified (Greiner 2000).   

To summarize, data from several galaxies, 
including our own, seem to indicate that
the numbers of SSSs are too small by at least two orders of magnitude to 
support the hypothesis that 
{\bf (1)}~the double-degenerate channel is the dominant way
to produce \t1e, and that {\bf (2)}~the winds incident on the first-formed \wdf\
 prior to the \ce\ phase
cause nuclear burning, and {\bf (3)}~the NBWDs can be detected and 
identified as SSSs.  
 
It is important to note, however, that the limits on the numbers of SSSs
that could correspond to \wdf s in the mass range corresponding 
to pre-double-degenerates are not as strong as the limits for the 
near-$M_C$ \wdf s 
expected in the single-degenerate model. Therefore more work is needed to
determine the fraction of SSSs, as a function of luminosity and temperature, 
that can be detected in each of several
nearby galaxies. Work that parallels the early calculations
(\rd\ \& Rappaport 1994) is needed.     
Specifically, the gas distributions of nearby galaxies can be modeled.
The galaxies themselves can then be seeded with SSSs, each with a given
luminosity and temperature. The numbers of counts expected from each
SSS during
observations with {\it Chandra} and {\it XMM-Newton}    
can then be computed to determine whether the source would have been
detected and, if so, whether there are enough photons to determine that
it is an SSS. In this way, we can determine the fraction of SSSs that can be
identified in each galaxy as a function of luminosity and temperature.
We can then discover how many sources are obscured by interstellar
absorption.  

\subsection{Implications of the Mismatch}

What does the lack of SSSs tell us
about the \pr s of \t1e? 
To answer this question we consider in turn each of the possibilities 
listed above.

\noindent {\bf (1)}~Perhaps the double-degenerate 
channel is {\sl not} 
the dominant way
to produce \t1e. 
In other words, the lack of SSSs is a sign that the NBWDs needed for
the \d-d\ model simply do not exist.   

We note here, however that
this conclusion
 does not follow simply
from the fact that there are too few SSSs. In fact, if we combine the   
result above for \dd s with the result of \rd\ 2010 for \sd s, and also 
assume that the lack of SSSs is due to the lack of NBWDs, then
all models of
\t1e\ involving \wdf s that accrete and process matter are eliminated.

\noindent {\bf (2)}~Perhaps the winds incident on the first-formed \wdf\
 prior to the \ce\ phase
do not cause nuclear burning. 
There are two possibilities: (a) the winds are not incident 
at the required rate, or  (b) the steady-burning regime does not
exist. 

(a) As Figure 1 demonstrates, the rate of wind infall is in or near
the steady-burning regime for a wide range of assumptions about winds,
the fraction of winds captured, and the exact boundaries of the steady-burning
regime. The only way that the wind infall could not be
adequate is if winds for the giant can be deflected by radiation and/or winds
from the \wdf .
In order for this to occur, however,
the \wdf\ must be generating a great deal of energy, perhaps
suggesting that nuclear burning must occur.  

(b) The existence of the nuclear-burning regime has been 
questioned (see e.g., Starrfield et al.\, 2005). 
Nevertheless, independent calculations by a number of groups find that
quasisteady nuclear burning is possible; furthermore there is
rough agreement on the locations of the upper and
lower boundaries of the steady burning regime (Iben 1982; Nomoto 1982; Shen
\& Bildsten 2008).
On general grounds, it seems likely that quasisteady nuclear burning
should occur over some range of infall conditions, since it should be
possible for accretion at high rates to produce conditions similar to
those found near the cores of giants.  
 
\noindent {\bf (3)}~Perhaps only a small fraction of
 NBWDs {\sl can} be detected and 
identified as SSSs. This seems to be the  
most likely possibility. First, circumstellar material can reprocess
the ultraviolet and soft-x-ray radiation, producing radiation at
longer wavelengths.  In symbiotics, the winds from the giant can
play this role. In fact symbiotic nebulae are common (see, e.g.,
Kenyon \& Murdin 2000). Second,
the duty cycle of nuclear burning can be low, as it is for recurrent novae.  
In the next section we consider a range of measurable
signatures for pre-\d-ds, such as orbital
period and stellar age.

\section{Symbiotics as Pre-Double-Degenerates}

We have shown that many double-degenerate binaries are descended  
from binaries in which the first-created \wdf\ accretes and burns
matter from a giant stellar companion. These binaries are
symbiotics. (See Kenyon \& Murdin 2000 and references therein.)  
It is possible that some of the 
known symbiotics are headed toward double-degenerate futures.
It will be important to identify such systems.
To do so, we must establish whether the giant will come to fill its 
\rl . If so, will the relative masses and rates of mass and angular momentum
loss be such as to trigger a common envelope? 
Finally, will the total \wdf\ mass exceed the Chandrasekhar mass?

The calculations that produced Figure 1 can be used to study the properties
of the pre-double-degenerates that are progenitors of \t1e. Each point in
Figure~3 represents such a binary in which the total white dwarf mass is 
greater than $M_C.$ Green (red) points correspond to systems in
which the \wdf\ gains more than $0.15\, M_\odot$ (less than $0.05\, M_\odot$).  
Extended periods of nuclear-burning may occur in most or all of these
systems. For example, $10^5$~years is required for a \wdf\ to gain $0.05\, M_\odot$
at an average rate of $5\times 10^{-7} M_\odot$~yr$^{-1}.$ Systems in which
more mass is gained, however, must burn material more quickly and
therefore be brighter, and/or must burn material over a longer period of time
and therefore be potentially detectable for a longer duration. 
The top panel of Figure~3 shows that the systems which gain more mass
tend to have larger orbital periods: longer than $5$~years for \wdf s gaining more than
$0.15\, M_\odot,$ although mass can be gained
by \wdf s in binaries with shorter orbital periods. 
The bottom panel shows that the longer orbital periods for systems in
which more mass is transferred are associated with the fact that the giant
donor in such systems becomes more evolved, achieving a larger 
core mass and radius before the \ce\ is triggered. Interestingly enough, these are the 
systems in which the two \wdf s that eventually merge will have a
total mass well over $M_C$.  

Also shown in Figure~3 is  
the lifetime, $\tau,$ of star ``2'', the star that became the giant donating mass 
to the \wdf . The value of $\tau$ is roughly equal to the time after the
formation of the binary when the nuclear burning activity occurs.  
From the perspective of detectability, we find that many progenitors will
experience periods of nuclear-burning activity
at times earlier than a few times $10^8$ years after formation; these systems
will be found near regions of star formation. In systems that gain the most mass, 
star ``2'' will generally have finished its epoch of mass transfer by
 $10^9$~ years after formation.
On the other hand, nuclear-burning activity may occur at late times in
binaries in which the \wdf\ gains less mass. Note that the value of $\tau$ is
also a measure of the initial mass of the secondary. 
Those \dd s in which the total \wdf\ mass will be largest are formed from
systems that experience an epoch of nuclear burning before roughly $10^8$~years;
both stars must start with masses larger than a few $M_\odot$; the orbital
period just prior to the \ce\ can be longer than $20$ years.  

Figure~4 shows the evolution of some individual binaries in our simulation
in which the total \wdf\ mass will exceed $M_C.$ We can use this figure to gain
additional insight into the possible appearance of the symbiotics that
are \pr s of \t1e through the \d-d\ channel.    
In this figure, $t=0$
would correspond to the time at which the \ce\ is triggered; $t$ is therefore
the time prior to the common envelope.  
The bottom panels show $\dot M_{in}$ versus $Log[t]$. 
The black curves correspond to the steady-burning region. The top panels show
the loss of the giant's mass through winds, the growth of its core mass, and the
growth of the mass of the accreting \wdf . All of the systems start with
such small winds that $\dot M_{in}$ is in the nova range; the white dwarf gains no mass
during the early stages of mass transfer. As winds increase, the \wdf s enter the
range of mass infall rates associated with recurrent novae. The system on the left
stays in this region for $\sim 10^6$~years; the \wdf\ gains only a small amount
of mass. The system in the middle panels passes through the recurrent nova region
in roughly $\sim 10^6$~years and then spends a comparable amount of time in the
steady-burning regime.
The mass increase of the \wdf\ is somewhat larger.
In the set of panels on the left, the mass infall rate increases beyond the 
steady-burning regime, and the system spends significant time in all three
mass-gain regions (recurrent novae; steady-burning; and in the region
above steady-burning, expected to be associated with heavier winds).

During the recurrent nova phase, the duty cycle of nuclear burning and of
potential detection as an SSS is small. For systems like RS Oph, 
it can be on the order of a percent (see, e.g. Nelson et al.\, 2009). 
Low duty cycle can help to explain the
small observed numbers of SSSs; systems like the one shown 
in the left-most panels
have a low probability of being detected in a nuclear-burning state.
On the other end of the scale, accretion at very high rates
can mean that the circumstellar region is dense, so that soft x-rays and
EUV radiation are absorbed and reprocessed. Systems like the one
shown in the right-hand panels may not be detectable as SSSs.

Figure~5, produced by the same simulation, shows the 
duration, $t_{active}$ of
nuclear-burning activity versus $\tau,$ 
the time at which the activity starts.
The systems in the top (bottom) panels produce \dd s with total \wdf\ mass 
greater than (less than) $M_C$. 
As in Figure~4, 
we consider separately the regimes (1)~below $\dot M_{min},$ where
recurrent novae are expected (panels on the left), 
(2)~the steady burning regime  (panels in the middle), and 
(3)~the regime of infall
rates greater than $\dot M_{max}$ (panels on the right).  
A difference between Figure~5 and Figure~4 is that each panel in 
Figure~4 shows the
evolution of an individual binary, while Figure 5 summarizes what happens
during the evolutions of all binaries in the simulation. In addition,
all binaries that enter the steady-burning regime start in the
recurrent novae regime.
Thus, each point in the middle panel {\sl also} corresponds to
a point in the panel to the left of it. Similarly,
each point in the ``heavy winds'' panels {\sl also} corresponds to   
a point in the steady-burning panel, and also to a point in the 
``recurrent novae'' panel. 
Figure~5 verifies that, for many systems,
 the duration of nuclear-burning activity is long,
in excess of $10^6$ years. It shows that systems in which
the mass accretion rate is above the steady-burning regime should end
their activity within a few times $10^8$ years after they are formed; this
is because the mass of the donor star tends to be large. This figure
also demonstrates that significant nuclear-burning activity is
expected from those pre-\d-ds that cannot become \t1e, because the
total \wdf\ mass is too low. White dwarfs in
these binaries can become active nuclear 
burners $\sim 10^{10}$ years after star formation. We may expect to
find them in elliptical galaxies and other old stellar populations.

The common 
envelope which ends the symbiotic epoch takes a relatively
short time, $\sim 10^5$~years to dissipate. The orbital separation at the
end of the common envelope phase determines the amount of time required
 for the \wdf s to come close enough to each other for mass transfer and merger
to occur. Depending on the efficiency of common 
envelope ejection in each binary, the interval between the nuclear-burning
phase and the supernova
could be shorter than $10^8$years, or longer than the Hubble time.

\section{Conclusion}

We have shown that many pre-double-degenerate binaries pass through an epoch
during which the first-formed \wdf\ 
accretes and burns matter from a giant companion.
If \dd s with total \wdf\ mass greater than $M_C$ comprise the major
component of \t1e \pr s, then the numbers of symbiotics with NBWDs must be
on the order of a thousand in galaxies such as our own. If the nuclear burning 
episodes produce SSS-like signatures, then 
we should be able to identify the pre-\d-d\ \pr s of \t1e in other galaxies
by identifying SSSs.  
In \S 4.1 we have
sketched the steps needed to determine the numbers of SSSs in external 
galaxies that have the luminosities and temperatures predicted for the pre-\d-d\ \pr s
of \t1e. Already, however, data
 from M31, M101, more than $380$ additional galaxies, 
and from the Milky Way strongly indicate that  
there is a mismatch
of $\sim 2$ orders of magnitude 
between the predicted numbers of SSSs 
and the numbers we actually detect in other galaxies. 
The result holds for young and old stellar populations.

We have already derived an even stronger result for 
single-degenerate \pr s of \t1e (\rd\ 2007; \rd\ et al.\, 2010; \rd\ 2010). 
We falsified the hypothesis that the
\s-d\ channel in which \wdf s accrete and burn enough matter to reach $M_C$
is the primary \pr\ channel and that the NBWDs  
appear as SSSs\footnote{ 
A similar result was claimed for old stellar populations 
based on limits on the diffuse soft  
emission from the Bulge of M31 and several early-type galaxies
Gifavov \& Bogd{\'a}n 2010). 
In these 
cases, however, 
the bright, hot NBWDs with masses near $M_C$ would have been 
detected directly had they been there, so the previously-existing 
limits apply.  
}
Combining the  
 results for \d-ds \s-ds, we find that 
there are not enough SSSs in our own and other
galaxies to explain the observed rates of \t1e. Since the supernovae occur,  
the implication is that there is a disconnect between either (1)~mass infall
at high rates and nuclear burning, or (2)~nuclear-burning
and SSSs. 

\noindent{\bf (1)} If mass 
infall doesn't lead to nuclear burning, this would seem
to imply that a change is needed in our understanding of fundamental
astrophysics. 
An alternative is that when nuclear-burning does occur, 
enough energy is released to deflect winds,  providing a kind of
thermostat mechanism. 

\noindent{\bf (2)} Mass accretion and 
nuclear burning is not always linked to SSS-like behavior.
This is already known to be the case for many systems. 
For example the duty cycle of SSS-like behavior is low for recurrent novae;
some of these, such as RS Ophiuchi are symbiotics (Nelson et al.\, 2009
and references therein).
In addition, absorption is expected because winds from symbiotics can absorb
radiation from the \wdf . In fact the nebulae associated with symbiotics
illustrate this point (see, e.g., Kenyon \& Murdin 2000).

With regard to the question of ``hiding'' the progenitors of \t1e,
symbiotics are intriguing for three reasons. First, of course,
is the likelihood that at least some symbiotics are
progenitors of \t1e. In this paper we have focused on
pre-\d-ds that may be \t1\ \pr s. Even among single-degenerates,
however, there are models in which the \pr\ passes through a phase in which a
giant donates mass to a \wdf\ either through winds or through Roche-lobe
overflow (\rd\ 1996).

Second, symbiotics are examples of very bright systems that have
proved difficult to identify.  
Estimates of the numbers of Galactic symbiotics are as high as
$4 \times 10^4$ (Magrini et al.\, 2003).  
In spite of these large numbers, and in spite of the fact that symbiotics
are, by their very natures highly luminous, the numbers of known symbiotics
had stood in the low hundreds until recently. Within the past several years, 
 $\sim 1000$ candidates,
now being checked, 
have been identified (see Corradi et al. 2009).
Whatever the appearance of the \pr s of \t1e, they too must be very bright,
at least during episodes of nuclear burning.  
They too, appear to be underrepresented, in that too-few candidates
have been identified in our Galaxy and in other galaxies.
Third, whether or not specific symbiotic binaries are \t1\ \pr s,
many contain NBWDs. The low numbers of SSSs we find is therefore relevant
for understanding the appearance of symbiotics.  

\smallskip

\noindent {\bf Summary:} 
The key issue identified by the study of galaxy populations of SSSs 
is that there are too few of them to serve as the \pr s of \t1e. This applies
to early-type and late-type galaxies. It applies to \s-d\ and \d-d\ models.
To understand the progenitors of \t1e, we must be able to predict the 
appearance of NBWDs and to identify a larger fraction of them in our 
own and other galaxies.

\bigskip
\noindent{\bf Acknowledgements:}  
It is a pleasure to acknowledge
helpful conversations, most recently with
 Scott Kenyon and Jeno Sokoloski, and also with
 participants
(especially Ed van den Heuvel, Lev Yungelson, and Jim Liebert),
 of the KITP conference
and workshop on {\it Accretion and Explosion} held at UC Santa Barbara in
2007. This work was supported in part by an LTSA grant from NASA
and by funding from the Smithsonian Institution.

\begin{figure*}
\begin{center}
\psfig{file=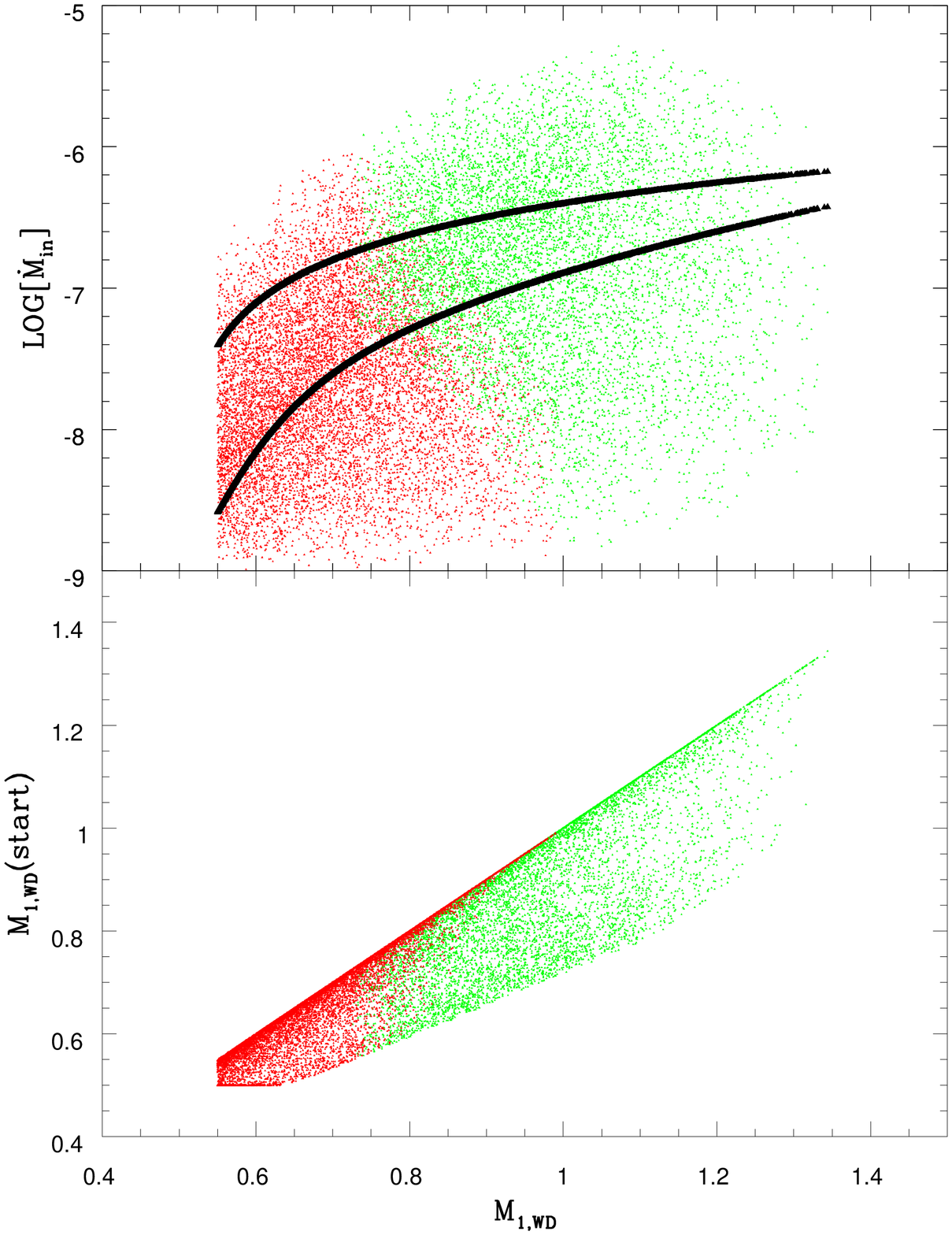,
height=7.0in,width=6.0in,angle=-0.0}
\vspace{-0 true in}
\caption{
Binaries in which a common envelope is about to be triggered. The \wdf, of
mass $M_{1,wd}$ orbits a more massive giant which is 
just about to fill its
\rl . 
{\bf Green (red) points} represent binaries in which the total white dwarf
mass will be larger (smaller) than $M_C.$ {\bf Top panel:}   
$0.25\, \dot M_{winds},$ the rate at which winds come under the
gravitational influence of the white dwarf, is plotted against the \wdf\ mass.
The two black curves show the lower and upper limits
of the steady-burning regime. {\bf Bottom panel:} the initial mass of
the \wdf\ is plotted against its mass just prior to the common envelope.   
}
\end{center}
\end{figure*}

\begin{figure*}
\begin{center}
\psfig{file=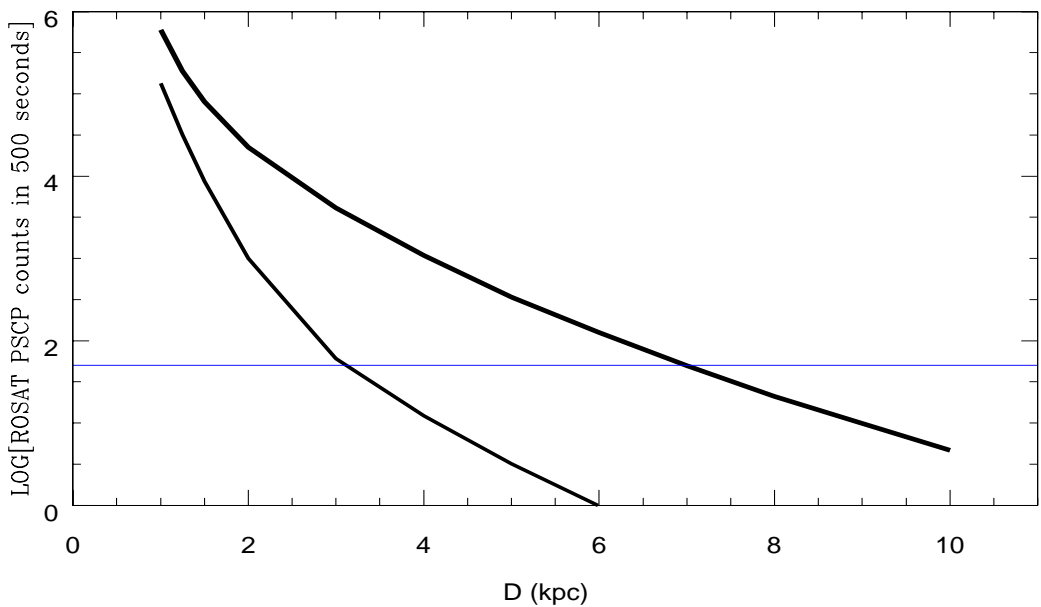,
height=7.0in,width=6.0in,angle=-0.0}
\vspace{-3 true in}
\caption{
Logarithm of the numbers of counts expected in a 500~s exposure with the
{\it ROSAT} PSPC, versus the source distance 
for a  
NBWD with mass: {\bf (top curve)}  $\sim 0.9\, M_\odot.$ 
We have taken $k\, T = 50$~eV
and $L = 3\times 10^{37}$~erg~s$^{-1}$.
{\bf (bottom curve)} $\sim 0.8\, M_\odot.$
We have taken $k\, T = 30$~eV
and $L =1 \times 10^{37}$~erg~s$^{-1}$. 
We have assumed a local density of one atom per cubic centimeter.
These curves therefore do not apply to star forming regions, nor to
directions toward the Galactic center. 
The horizontal line corresponds to $50$ counts; with this number of counts the
source would be detected and recognized as having a cut-off near or below $1$~keV. 
}
\end{center}
\end{figure*}

\begin{figure*}
\begin{center}
\psfig{file=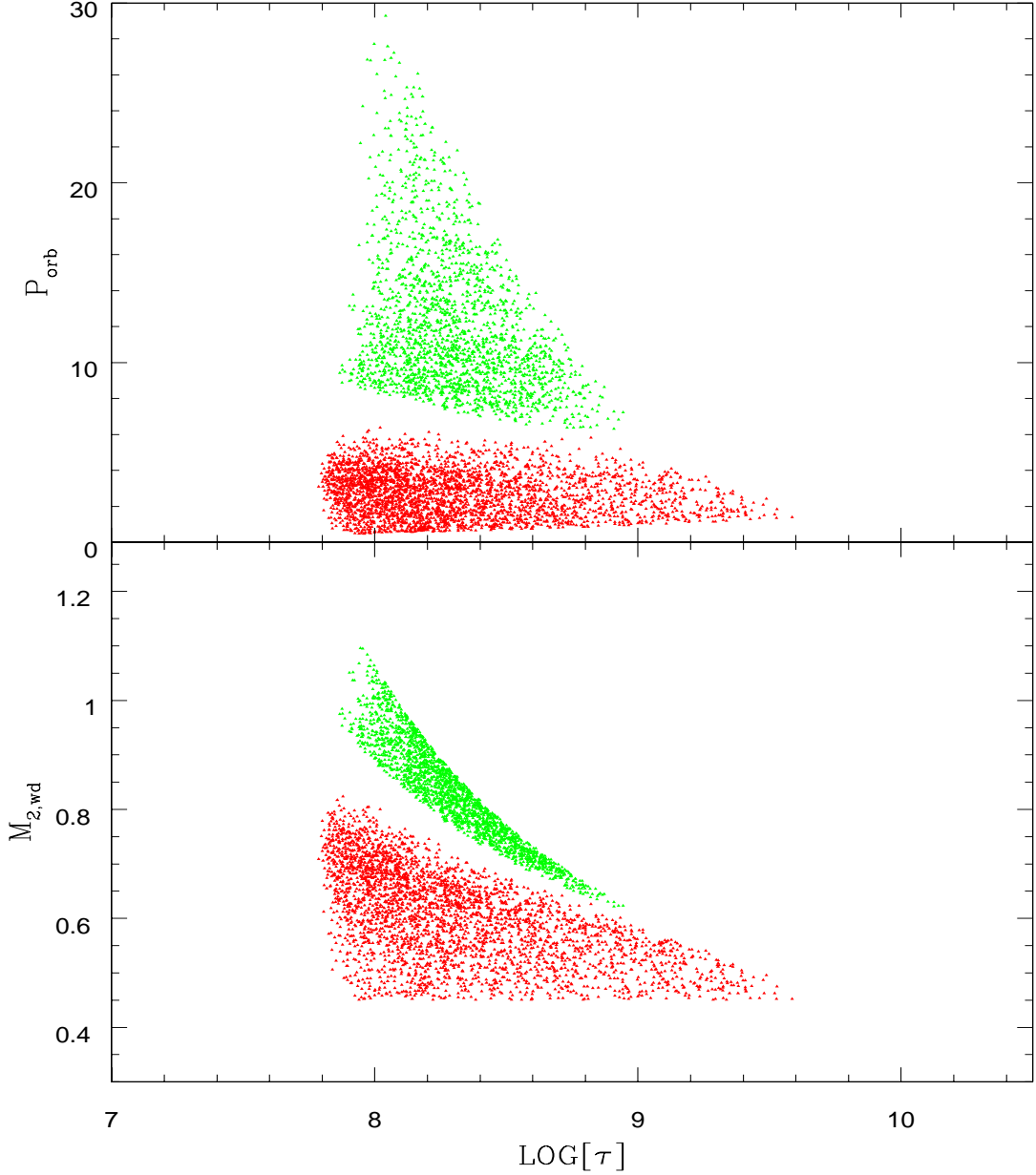,
height=7.0in,width=6.0in,angle=-0.0}
\vspace{0 true in}
\caption{
Binaries in which a common envelope is about to be triggered. 
Each point corresponds to a system which is a candidate for an SNIa via the
double-degenerate channel: $M_{1,wd}+M_{2,wd} > M_C.$ {\bf Points in green:}~$M_{1,wd}$
has increased by more than $0.15\, M_\odot$ due to winds. {\bf Points in red:}~$M_{1,wd}$
has increased by less than $0.05\, M_\odot$ due to winds. 
{\bf Top panel:}~$P_{orb}$ versus the lifetime of the secondary, $\tau.$  
{\bf Bottom panel:}~$M_{2,wd}$ versus the lifetime of the secondary, $\tau.$  
}
\end{center}
\end{figure*}

\begin{figure*}
\begin{center}
\psfig{file=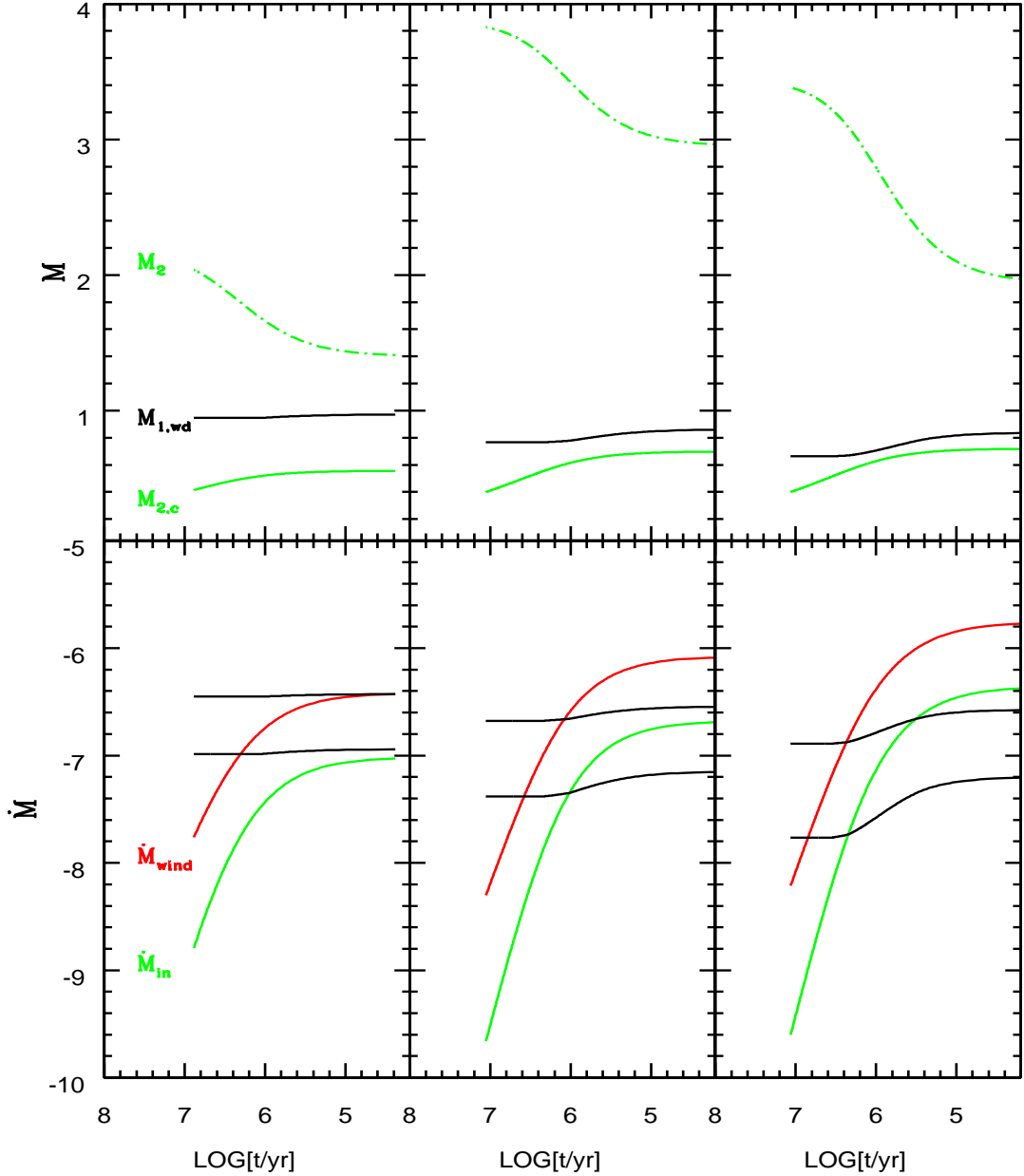,
height=7.0in,width=6.0in,angle=-0.0}
\vspace{0 true in}
\caption{
The evolution of binaries which will experience a common envelope, with
$M_{1,wd}+M_{2,wd}>1.4\, M_\odot.$  
{\bf Top panels:}~Mass evolution, with green curves showing the evolution of the
giant's total mass (dashed, dotted) and its core mass (solid), and black showing the
mass of the white dwarf. 
{\bf Bottom panels:}~Evolution of the mass flow, with red corresponding to $\dot M_{wind}$
and green to $\dot M_{in}$. The black curves show the minimum and maximum rates
compatible with quasi-steady nuclear burning, $\dot M_{min}$ and
$\dot M_{max}.$   
Plotted along the horizontal axis is the logarithm of the time in years, prior to
the common envelope.  
}
\end{center}
\end{figure*}

\begin{figure*}
\begin{center}
\psfig{file=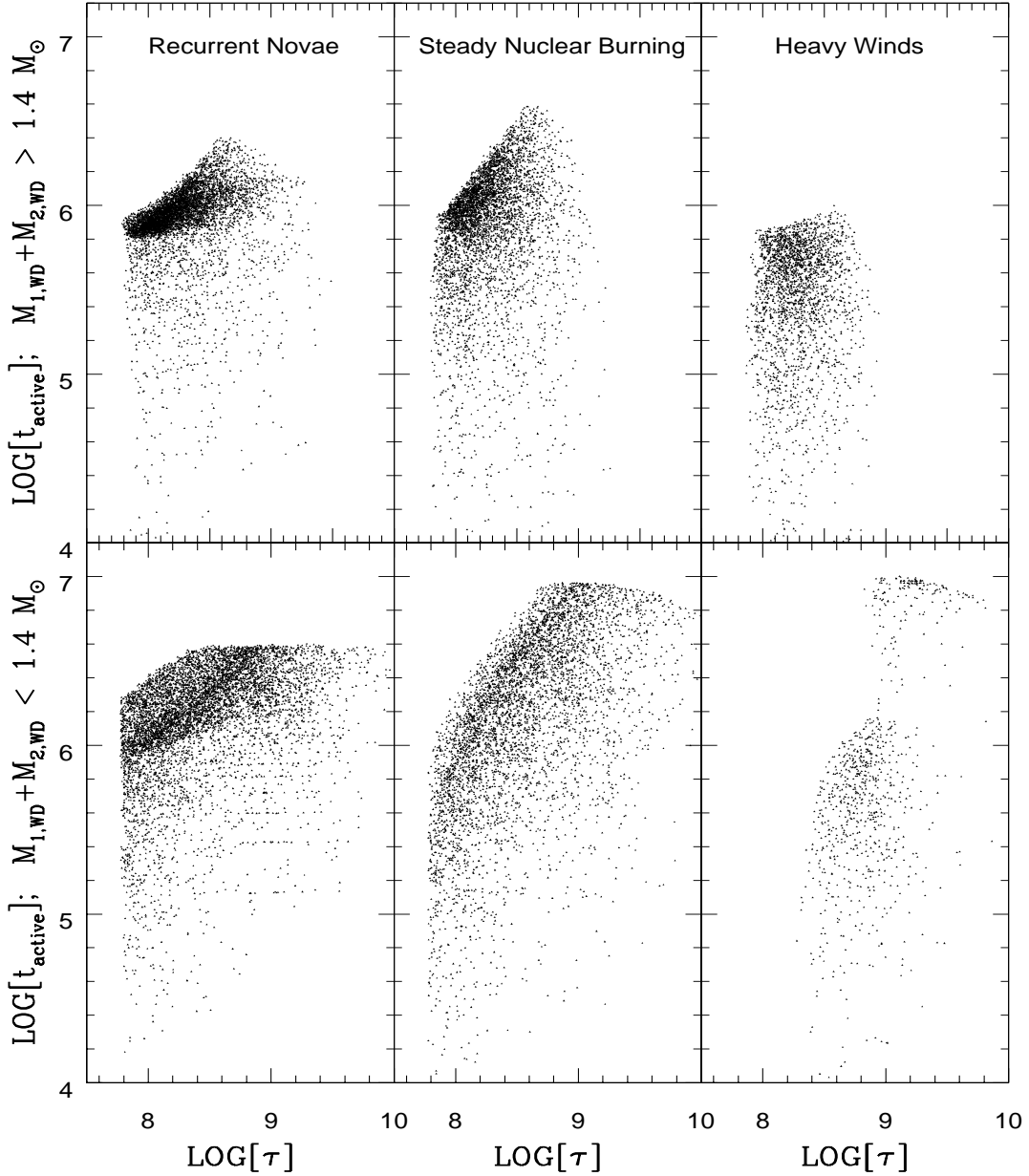,
height=7.0in,width=6.0in,angle=-0.0}
\vspace{0 true in}
\caption{
Logarithm of the duration of nuclear burning versus logarithm of the start time.
{\bf Top panels:} Candidates for 
Type~Ia progenitors ($M_{1,wd}+M_{2,wd}>M_C$).  
{\bf Bottom panels:} The total \wdf\ mass is smaller than $M_C.$
{\bf Left-most panels:} $\dot M_{in}$ lies between $1/3\, \dot M_{min}$
and $\dot M_{min}$; nuclear burning occurs during recurrent novae. 
{\bf Middle panels:} $\dot M_{min} < \dot M_{in}< \dot M_{max}$; the
\wdf\ is most likely to be detected as an SSS. 
{\bf Right-most panels:} $\dot M_{in}>\dot M_{max}$; mass in winds may
absorb radiation emitted by the \wdf. 
}
\end{center}
\end{figure*}

\end{document}